\documentclass[a4paper,fleqn]{cas-sc}
\usepackage[numbers]{natbib}
\usepackage{hyperref}
\usepackage[dvipsnames]{xcolor}

\def\tsc#1{\csdef{#1}{\textsc{\lowercase{#1}}\xspace}}
\tsc{WGM}
\tsc{QE}
\tsc{EP}
\tsc{PMS}
\tsc{BEC}
\tsc{DE}

\DeclareMathOperator{\erf}{erf}
\begin{document}
\let\WriteBookmarks\relax
\def\floatpagepagefraction{1}
\def\textpagefraction{.001}

\shorttitle{Mass media and its impact on opinion dynamics of the nonlinear \textit{q}-voter model}

\shortauthors{R.~Muslim, R.~Anugraha~NQZ, and M.~A.~Khalif}

\title[mode = title]{Mass media and its impact on opinion dynamics of the nonlinear \textit{q}-voter model}    
\author[1]{Roni Muslim}[orcid= 0000-0001-6925-5923]
\ead{roni.muslim@brin.go.id}

\author[2]{Rinto Anugraha NQZ}[orcid=0000-0002-4781-4090]
\ead{rinto@ugm.ac.id}
\cormark[1]
\cortext[cor1]{Corresponding author}
\author[3]{Muhammad Ardhi Khalif}[orcid=0000-0002-7807-1496]
\ead{muhammad_ardhi@walisongo.ac.id}

\affiliation[1]{organization={Research Center for Quantum Physics,
    National Research and Innovation Agency (BRIN)},
    city={South Tangerang},
    postcode={15314}, 
    country={Indonesia}}

\affiliation[2]{organization={Department of Physics,
    Gadjah Mada University},
    city={Yogyakarta},
    postcode={55281}, 
    country={Indonesia}}

\affiliation[3]{organization={Department of Physics,Walisongo State Islamic University},
    city={Semarang},
    postcode={50181}, 
    country={Indonesia}}

\begin{abstract}
With the success of general conceptual frameworks of statistical physics, many scholars have tried to apply these concepts to other interdisciplinary fields, such as socio-politics, economics, biology, medicine, and many more. In this work, we study the effect of mass media on opinion evolution based on the nonlinear $q$-voter by means with probability $p$ a voter adopts the mass media opinion whenever a $q$-sized agent in the population is not in unanimous agreement. We perform analytical and numerical calculations for some quantities of macroscopic parameters of the model such as order parameter (representing an average of public opinion), consensus (relaxation) time, and exit probability, and obtain the agreement results. We find the power-law relations for some quantities of the model. (1) The probability threshold $p_t$, i.e a probability that makes the system reaches a homogeneous state, follows the power-law relation $p_t \sim q^{\gamma}$ with the $q$-sized agent, where $\gamma = -1.00 \pm 0.01$ is the best fitting parameter. The probability threshold $p_t$ also eliminates the {coexistence two ordered states} of the model. (2) The relaxation time (the time needed by the system to reach consensus)  $\tau$ with the population size $N$ is obtained in the form of $\tau \sim N^{\delta}$, where $\delta$ depends on the probability $p$ and $q$-sized agent. We also approximate the {separator} point {$r_s$} and the system's scaling parameters by employing the standard finite-size scaling relation. 
\end{abstract}




\begin{keywords}
Opinion dynamics \sep exit probability \sep relaxation time \sep scaling \sep mass media effect.
\end{keywords}

\maketitle

\section{Introduction}
The researchers translate the simple interaction in the Ising spin model into real social life by making a limited interaction rule between individuals that is still in line with the rules of the Ising model. These rules of social interaction are referred to as the model of opinion dynamics. This model is a rule of opinion formation and change due to the interaction between individuals and can be formulated as a mathematical equation \cite{lorenz2007continuous,castellano2009statistical,noorazar2020classical, zha2020opinion, qesmi2021dynamics}. The interaction between individuals makes individuals similar in terms of, for example, thought and behavior. This kind of interaction is identical to the behavior of the spins which interact with each other in the Ising model, where the stability condition due to interacting with neighboring spins causes a change in state, where the spins will flip to a more stable state than the previous condition forming a parallel state \cite{ising1924beitrag}. Moreover, concepts in statistical physics are used to describe social phenomena such as hysteresis, agreement or consensus, disagreement or status quo indicated by the existence of the order-disorder phase transition \cite{rogers2013consensus,oestereich2020hysteresis,neuhauser2020multibody,krapivsky2021divergence, papanikolaou2022consensus, MUSLIM2022133379, MUSLIM2022128307}. These social phenomena can be analyzed from macroscopic parameters such as magnetization that describe the average opinion of agents in the system.

In the opinion dynamics model, one can observe some properties of statistical physics, such as phase transitions, scaling, and universality. Such statistical physics properties usually arise because of the ``insertion'' of noise parameters to the model, which in the social context refers to individual behavior that is always contrary to the prevailing norms in society, such as nonconformity behaviors \cite{javarone2014social, krueger2017conformity, muslim2020phase, muslim2021phase, MUSLIM2022133379, MUSLIM2022128307}. Such individual behaviors undermine group agreement and can lead to a system undergoing a stalemate state or status quo. Another interesting phenomenon in nature is that the scaling behavior of some phenomena which can be observed so far, such as dynamics of an economic index \cite{mantegna1995scaling},  the voting distribution in the Brazilian election in 1998 has a scaling parameter $\approx -1.0$ \cite{costa1999scaling}, the human locomotor activity \cite{indic2011scaling}, the quantum phase transition \cite{wu2020scaling}, the stiffness and strength of hierarchical network nanomaterials \cite{shi2021scaling},  and many more.

One of the interesting scenarios to study in an opinion dynamics model is how the mass media affects the change of opinion in each individual and its impact on the evolution of opinion on a large scale, namely in the community. A previous study has shown that the effects of the mass media are very influential in an individual's opinion. Individuals can change or keep their opinions on a matter due to the influence of the mass media \cite{pinkleton1998relationships}. Other previous works have reported the effect of mass media on various models of opinion dynamics with various scenarios, which can be seen in Refs. \cite{mazzitello2007effects,candia2008mass,rodriguez2010effects,pabjan2008model,sousa2008effects, crokidakis2012effects,gonzalez2007information,gonzalez2005nonequilibrium,martins2010mass,quattrociocchi2014opinion,pineda2015mass,colaiori2015interplay,sirbu2017opinion,li2020effect,freitas2020imperfect,tiwari2021modeling,mus2020,civitarese2021external}. The mass media or sometimes referred to as the external field in some studies has shown various effects on the evolution of opinion, depending on the considered scenario. It has been reported that the absence of mass media can speed up the process of consensus on diversity \cite{pineda2015mass}. Another interesting result reported that the presence of the mass media could shift the system from pluralism to consensus or agreement in terms of public opinion \cite{colaiori2015interplay}. Another recent study showed how the initially heterogeneous public opinion could reach a fixed point if introduced by the mass media \cite{gimenez2021opinion}.

As mentioned above, the interesting situation in real life is how the existence of external influence affects opinion formation. The concept of external influence is analogous to the external magnetic field in the Ising model, i.e., it can change the microscopic and or macroscopic behavior of the system. The external field in this model can be considered as the mass media in real life setting, i.e., an individual will follow the mass media opinion with probability $p$. Nowadays, it is well known that the advances in information technology make it easier for individuals to connect through the media. Therefore, individuals likely obtain information from the mass media and influence their opinions. The mass media can easily affect individuals, resulting in a shift in their opinion. In the Ising model context, mass media's effect can be represented by its impact on the coexistence of the ordered states. Previous work reported that the presence of mass media could eliminate the {coexistence two ordered states} (called as usual phase transition by author) of the model. At a certain probability $p$, the effect is that all individuals, in the end, share the same opinion (all parallel up) for a small value of the initial fraction of ``up'' opinion \cite{crokidakis2012effects}. However, the work only presented numerical simulation results and examined several macroscopic states of the system such as the order parameter and relaxation times.

This paper studies the nonlinear $q$-voter model, defined on a fully connected network or complete graph. A fully connected network is a structure where links or edges connect all nodes. The nodes represent individuals or agents has two possible opinions, as Ising number $\pm 1$. We perform analytical calculations and numerical simulations to analyze the macroscopic behavior of the system. The effect of mass media is similar to the scenario implemented in the two-dimensional Sznajd model on a square lattice. A voter follows the mass media opinion with probability $p$ if the $q$-sized agent is not shared the same opinion \cite{crokidakis2012effects}. We find that the {coexistence of two ordered states} is absent at a probability threshold $p_t$ that is measured at a small value of initial ``up'' opinions at the system. The system reaches a complete consensus at that probability threshold with the order parameter $m = 1$. Results from numerical Monte Carlo (MC) simulation agree very well with the analytical calculations. The probability threshold $p_t$ follows a power-law relation to the $q$-sized agent, $p_t \sim q^{\gamma}$, where $\gamma \approx -1.00$ is the best-fit parameter value for the data. The value of $\gamma$ is similar to the scaling parameter reported on the voting distribution among candidates in the Brazilian election in 1998 \cite{costa1999scaling}. We also analyze the {separator} point of the model and scaling parameters using finite-size relations and find the robust scaling parameters for a range of $q$-sized agent ($\approx [2-100]$). The {separator} point also called the unstable point of the system, allows the {existence of two ordered states}. 

Moreover, the {separator} point will separate the ferromagnetic state (analogous to complete consensus) with $m = -1$ and $m = +1$. In addition, we also examine the model's relaxation time (consensus time), which is the duration required for the system to reach a complete consensus from a disordered state. The probability $p$ decreases the system's relaxation time as $p$ increases for the same population size $N$. The relaxation time parameter $\tau$ also obeys a power-law relation to the population size $N$, which can be expressed as $\tau \sim N^{\delta}$, where $\delta$ depends on the probability $p$ and $q$-sized agent. The exit probability of the system, i.e., the probability of the system being in a state of complete consensus with an ``up'' opinion, collapse with scaling parameters $\alpha_1 \approx 0$ and $\alpha_2 \approx 0.5$ for all values of population $N$.

\section{Model and methods}
\label{sec.2}
To analyze the opinion evolution on the system, we consider the order parameter of the system that is defined as:
\begin{equation}\label{eq:Eq01}
    m = \frac{\sum_{i}^{N}\sigma_i}{N},
\end{equation}
where $N$ is the total population in the system (total number of sites), which is defined as $N = N_+ + N_-$. $N_+$ and $N_-$ are the total agents with state or opinion ``up'' and ``down'', respectively. The variable $\sigma_i = \pm 1$ is spin Ising represents that each agent has two possible states or opinions. The order parameter $m$ represents the average of the public opinion per site, where for $|m| = 1$, indicating the public opinion is in complete consensus, for $|m| < 1 $, there is a majority or minority opinion in the system, and for $m = 0$, the system is in a stalemate situation or status quo. In statistical physics, the complete consensus situation is similar to the ferromagnetic state, while the status quo is similar to the antiferromagnetic state. 

The ``nonlinear'' term in the nonlinear $q$-voter model refers to the nonlinear form of the flip function below \cite{dornic2001critical,vazquez2008systems}:
\begin{equation}\label{eq:update}
  f = z^q + \varepsilon \left[1-z^q-(1-z)^q\right],
\end{equation}
where $z,q,$ and $\varepsilon$ are the fraction of opposite neighbors, the size of the interaction agent, and the noise parameter. We can check that for $q = 1$, the model becomes the linear $q$-voter model, $f = z$. In addition, we can obtain the linear form of Eq.~\eqref{eq:update} for a specific case $q = 2$ and $\varepsilon = 1/2$. In general, the nonlinear form of Eq.~\eqref{eq:update} ($f \neq z)$ can be obtained for $q \neq 1$  \cite{castellano2009nonlinear, moretti2013mean}. In the nonlinear $q$-voter model, each voter follows the opinion of the $q$-sized agent if that group has the same opinion. If the $q$-sized agent is not in an ordered state (not sharing the same opinion), the voter can still adopt the opinion with probability $\varepsilon$ \cite{castellano2009nonlinear}. This paper considers $ \varepsilon = 0$ as in Refs.~\cite{civitarese2021external,nyczka2012phase} and introduces a probability of an external effect $p$ to the model. Following the scenario by Crokidakis \cite{crokidakis2012effects}, we extend the nonlinear $q$-voter model by adding the effect of mass media on the fully connected network, which can potentially lead the system to reach the system voter to adopt the mass media opinion.

The algorithm of the nonlinear $q$-voter model with mass media effect can be described as follows:
\begin{enumerate}
    \item {Initially, the probability of opinion (state) up and down are $r$ and $1-r$, respectively}
    \item A $q$-sized agent and a voter in the population are picked randomly.
    \item If the $q$-sized agent shares the same opinion or state, the voter ($S_{q+1}$) follows the opinion of the $q$-sized agent, $S_q = S_{q+1}$. Otherwise,  with probability $p$, the voter adopts an opinion expressed by the mass media.
\end{enumerate}

\begin{figure}[t]
\centering
\includegraphics[width = 8 cm]{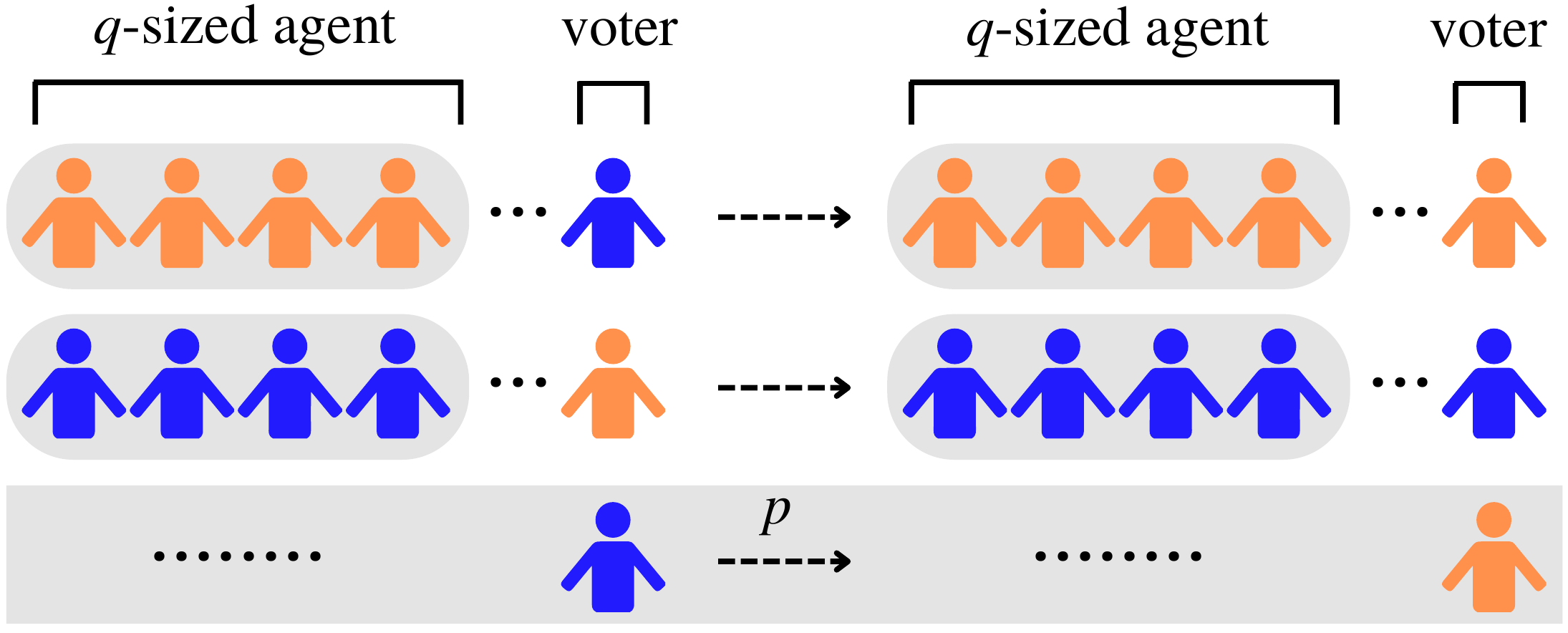}
\caption{\label{fig:Fig01} Illustration of the nonlinear $q$-voter model with $q=4$ ($q$-sized agent). Two scenarios at the top follow the standard nonlinear $q$-voter model, where the voter follows the $q$-sized agent's opinion if the $q$-sized agent is ordered. The bottom scenario shows how the voter changes their opinion from one state (blue color) to another state (orange color) with probability $p$ when the two top scenarios are not fulfilled. The orange ($+1$) and blue ($-1$) colors represent two different opinions or attitudes of each agent.}
\end{figure}

This model illustrates a social situation where the agent's $q$-sized opinion can represent the opinions of the ``closest people'', such as family, relatives, and close friends, whose opinion influence is strong and takes precedence for consideration. However, if the opinion of these close people is not unanimous, then the voter will {adopt the mass media opinion}, which is represented by the probability $p$. {The mass media leads the voter to choose one of two candidates, for example, by adopting opinion up (+1)}. The illustration of the model, for a case $q = 4$, is exhibited in Fig.~\ref{fig:Fig01}. As shown in Fig.~\ref{fig:Fig01}, we set the probability that the voter changes their opinion as $1$ in the case of the voter following the $q$-sized agent opinion and with probability $p$ in the case of following the mass media opinion {($-1$ to $+1$)}. Therefore the control parameter in this model is only the probability $p$ (bottom scenario).

We define the model on the fully connected network, where each agent (represented by a node) is linked to each other forming a complete connection. In the social system, the complete graph can be considered a group of people who can interact with each other anytime and anywhere, for example, by phone and online chat.  {In general, the representativeness of the network structure depends on the model being analyzed. For limited systems, such as small communities where everyone can interact directly, the complete graph is good enough to represent the system \cite{albert2002statistical, newman2018networks}. We consider the complete graph because the analytical treatment can be carried out conveniently. In addition, in the complete graph, we can find some common properties of statistical physics, such as the scaling parameters, phase transition, relaxation time, exit probability, and universality of the model}. In this work, we consider a mass media effect on the system by introducing the probability $p$ to the model. Each voter follows the mass media opinion whenever the $q$-sized agent is not shared the same opinion.

We will show that the probability $p$ affects the {separator} point of the system. To analyze the dependence of the phase transition of the system on the possibility $p$, we variate the probability of mass media $p$ given the small value of initial density of spin-up $r = 0.01$ instead of variating the density $r$ of spin-up states for a given probability $p$ of mass media as in Refs.~\cite{crokidakis2012effects, azhari2022mass}. We will analyze what value of $p$ makes all voters share the same opinion with the magnetization $m=+1$. The {coexistence of two ordered states} of the system will be eliminated by the presence of the probability $p = p_t$, i.e., we shall observe that at $p \geq p_t$, an ordered state or complete consensus has been reached by the system with magnetization $m = +1$. The difference value of the $q$-sized agent will give the difference of the probability threshold $p_t$, and we will find the relationship between the $q$-sized agent to the probability threshold $p_t$. We also investigate the relaxation or consensus time of the model, i.e., the time required to reach an ordered state. The relaxation time of the voter model has been investigated in a $d$-dimensional lattice following the relation $\tau \sim N^2$ for $d = 1$, and $\tau \sim N \log N$ for $d = 2$. In addition, the consensus time $\tau$ follows the relation $\tau \sim N$ for $d > 3$ \cite{blythe2010ordering} and also in the mean-field result \cite{krapivsky2010kinetic}. In this paper, we also find that the consensus time of the model follows the power-law relation $\tau \sim N^{\delta}$, where $\delta$ depends depends on the probability $p$ and $q$-sized agent.

\section{Results and discussion}
\label{sec.3}

\subsection{Probability threshold and {separator point}}
\begin{figure}[t]
\centering
\includegraphics[width = 8.5cm]{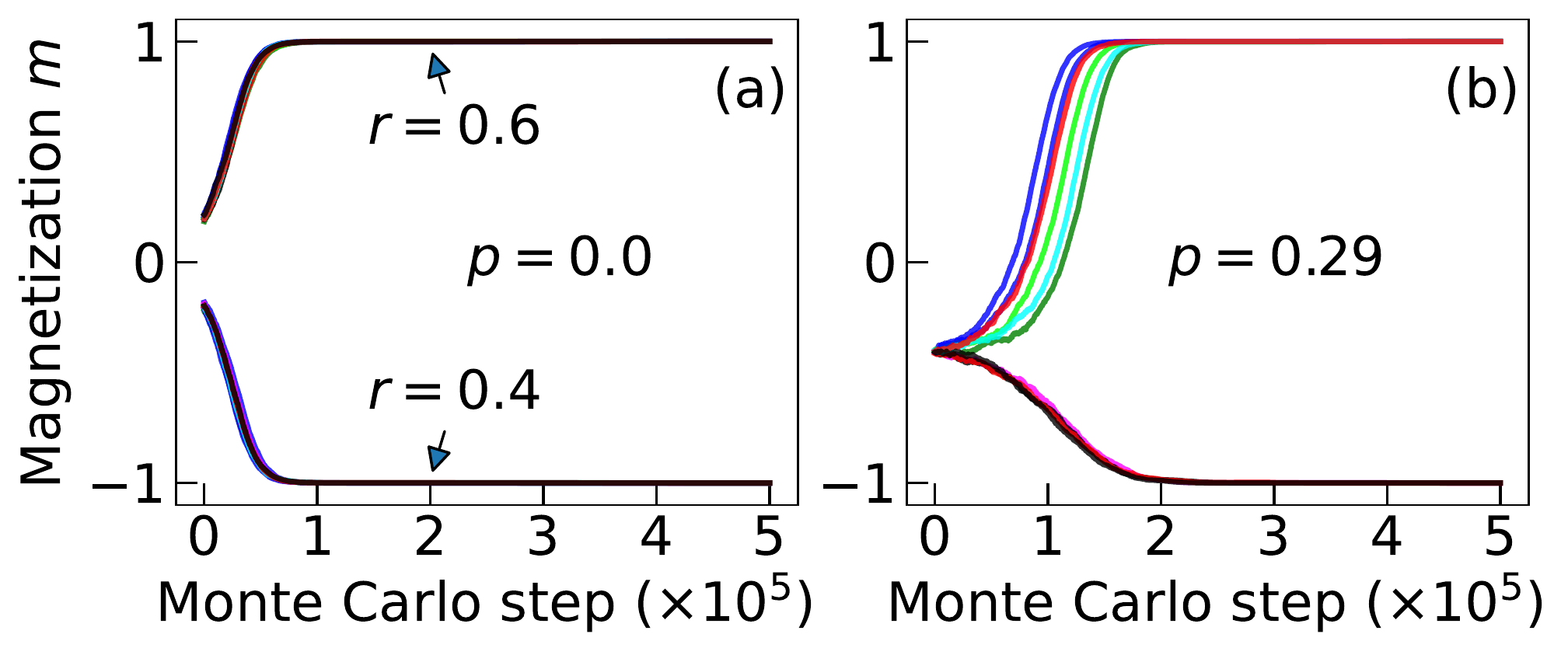}
\caption{\label{fig:Fig02} The evolution of the public opinion per site (represented by magnetization $m$) is based on the nonlinear $q$-voter model for $q = 2$. [panel (a)] Without the mass media effect ($p = 0$) for the typical spin-up densities $r_0 = 0.4 < r_s$ and $r_0 = 0.6 > r_s$. One can see that all samples reach an ordered state (complete consensus) with $m = -1$ for $r = 0.4$ and $m = 1$ for $r = 0.6$, characterized by the {separator} point at $r_s = 0.5$. [panel (b)] The case with the mass media effect with $p = 0.29$. One can see that even for the initial $r = 0.3 < 0.5$, all samples are not towards an ordered state with $m = -1$ as in the original model, indicating the {separator} point $r_s$ depends on the mass media effect $p$. The population size $N = 10^4$. }
\end{figure}

We find that for the case without mass media effect where $p=0$, the value of $r=0.5$ produces the unstable point of the system, which separates two ordered states $m = +1$ (all spins in parallel up) and $m = -1$ (all spins in parallel down) for spin-up density $r_0 > 0.5$ and  $r_0 < 0.5$, { respectively for a large system as shown in Fig.~\ref{fig:Fig02}~(a)}. The unstable point ${r_s} = 0.5 $  can be referred to as a {separator} point of the model. The original Sznajd model in the two-dimensional form (the case which does not include mass media effect) produced a similar result \cite{stauffer2000generalization}. We can say that when there is no probability $p$, all agents will compete neutrally to achieve an ordered state. Of course, {in the system with the large population, the large initial population will win}. {The existence of} the mass media effect, represented by a probability $p$, makes the model have a {separator} point no longer at $0.5$, {but instead shifted to another $r$ (somewhere) as shown in Fig.~\ref{fig:Fig02}~(b)}. It can be seen in Fig.~\ref{fig:Fig02}~(b), for the initial $r_0 = 0.3 < {r_s}$ and $p = 0.29 $, some samples show that the system has reached the complete consensus with $m=+1$. The result thus indicates that the {separator} point has shifted from $r = 0.5$ and depends on the probability $p$.

\begin{figure}[t]
\centering
\includegraphics[width = 6.5 cm]{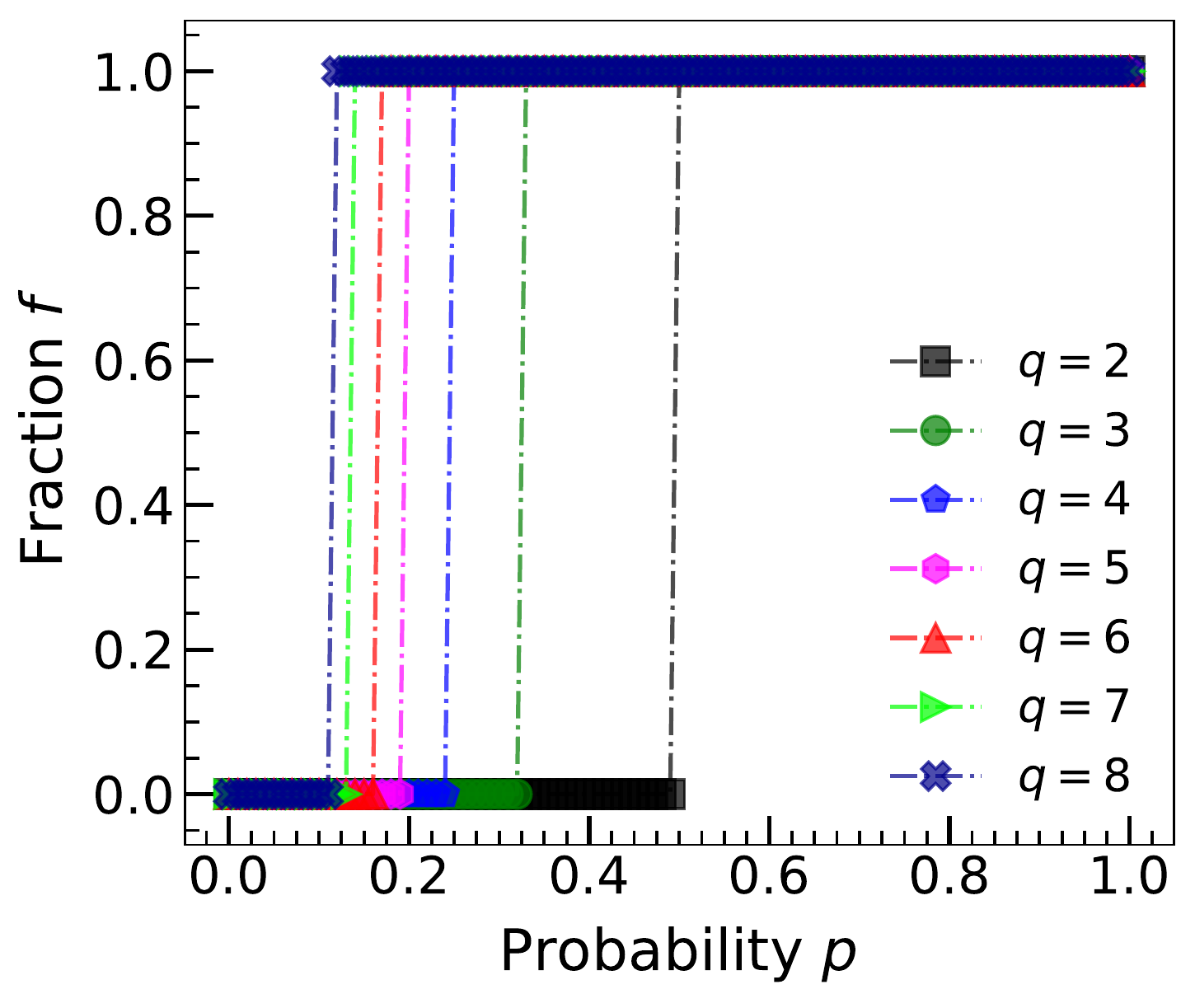}
\caption{\label{fig:Fig03}Fraction $f$ versus probability of mass media effect $p$ of the nonlinear $q$-voter model on the fully-connected network for a small initial value of spin-up density $r = 0.01$ and several values of $q$-sized agent, $q=2,3,4,5,6,7$ and $8$. For each $q$-sized agent, at a probability threshold $p =p_t$, the fraction $f$ `jumps' from $0$ to $1$, indicating that the {coexistence of two ordered states} has been eliminated. In other words, the {coexistence of two ordered states} is absent for $p \geq p_t$, i.e., for $p \geq p_t$, the evolution of the order parameter $m$ will go to the final state with all agents having the same opinions `up'. The population size $N = 10^5$.}
\end{figure}

To analyze the effect of the mass media on the opinion evolution of the system, we first perform a Monte Carlo simulation to compute the average of the order parameter $m$ to the change of the probability $p$.
We consider that if there {is a coexistence of two ordered states}
i.e., for a small initial value of spin-up density $r$, we will find that the average of the order parameter $m$ is always less than $+1$. In other words, the system reaches the ordered states with $m =+1$ and $m = -1$. Moreover, if, in the end, all spins in the state up ($m = +1$) at a probability of mass media $p$, then this situation indicates the absence of the coexistence of two ordered states.
Therefore, the mass media effect $p$ eliminates the {coexistence of two ordered states} at the probability threshold $p = p_t$. To analyze such phenomena, we consider a fraction parameter $f$, which represents an occurrence probability of the agents having an up opinion. Therefore, the fraction $f$ has a value range $0-1$, where $0$ and $1$ stand for all agents in complete consensus with $m = -1$ and $m =  1$, respectively. We fix an initial of the spin-up density $r = 0.01$ and variate the probability of mass media $p$ for typical values of $q$-sized agent. To obtain a good result, we utilize a large population size $N = 10^5$; each data point averages over $500$ samples, and $\Delta p = 0.01$. The Monte Carlo simulation result is exhibited in Fig.~\ref{fig:Fig03}. One can see a value of the probability threshold $ p = p_t$ (for each $q$-sized agent), which shows that all agents at the end have the same opinion (complete consensus), represented by the fraction  $f =  1$. The fraction value $f$ 'jumps' from $f = 0$ to $f = 1$ at probability threshold $p_t$ indicates the absence of the {coexistence of two ordered states} at $p \geq p_t$. In other words, the probability $p\geq p_t$ eliminates the {coexistence of two ordered states} in the system. There is a {coexistence of two ordered states} for $p < p_t$ for all $q$-sized agents, where the value of $p_t$ depends on the $q$-sized agent.

\begin{figure}[t]
\centering
\includegraphics[width = 6.5 cm]{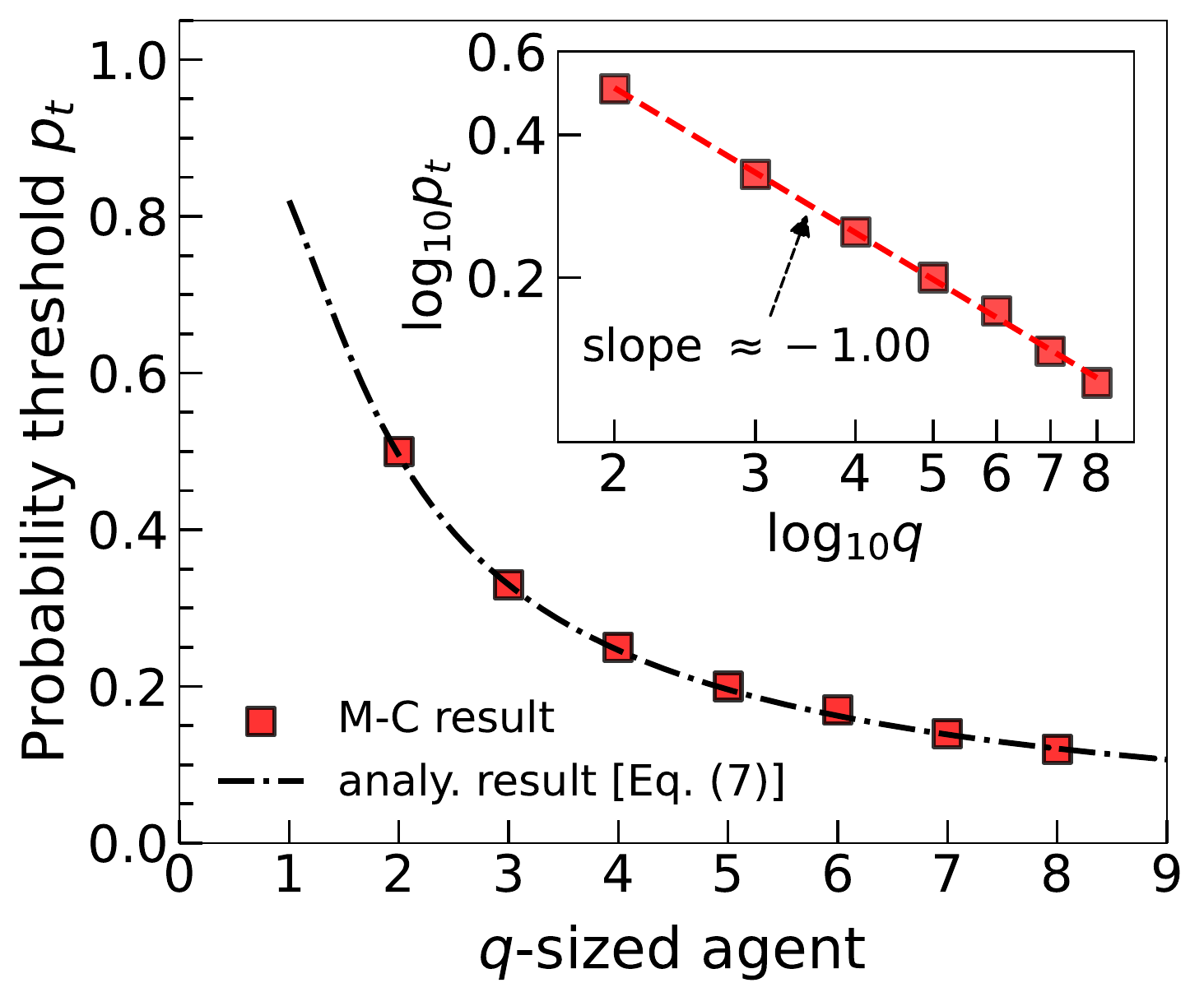}
\caption{\label{fig:Fig04} (Main graph) Comparison of the Monte Carlo simulation with the analytical results in Eq.~\eqref{eq:critical} for several values of $q = 2,3,4,5,6,7$ and $8$ (red data points). The M-C simulation agrees with the analytical result  (dashed line). The inset graph shows the log-log plot of the data, and the model follows the power-law relation $p_t \sim q^{\gamma}$, where $\gamma = -1.00 \pm 0.01$ is the best fitting parameter to plot the data. The scaling parameter $\gamma$ is similar to the scaling parameter obtained in the voting distribution in the 1998 Brazilian election \cite{costa1999scaling}. We utilize a large population size $N = 10^5$. }
\end{figure}

The probability threshold $p_t$ for all $q$-sized agents are $p_t \approx 0.50 \, (q = 2), 0.33 \, (q = 3), 0.25 \, (q = 4),  0.20 \, ( q = 5), 0.17 \, (q = 6), 0.14 \,(q = 7), 0.12\,(q = 8)$, respectively. Plot of $p_t$ versus $q$ is exhibited in Fig.~\ref{fig:Fig04}. One can see the agreement result between the Monte Carlo simulation (data point) versus the analytical calculation (discussed in the next section), represented by the dashed line. The exact value of the probability threshold $p_t$ is relatively difficult to determine numerically. However, we can confirm this result through further analytical and other numerical calculations. Furthermore, based on our numerical simulations, we find the probability threshold $p_t$ follows a power-law relation to the $q$-sized agent, that is $p_t \sim q^{\gamma}$, where $\gamma = -1.00 \pm 0.01$ is the best scaling parameter of the data as exhibited in the inset of Fig.~\ref{fig:Fig04}. This scaling behavior is similar to that in the 1998 Brazilian election reported by Costa et al. \cite{costa1999scaling}. We have checked that the value of $\gamma$ is valid for $2 \leq q  < 100$.

\begin{figure}[tb]
\centering
\includegraphics[width = 16 cm]{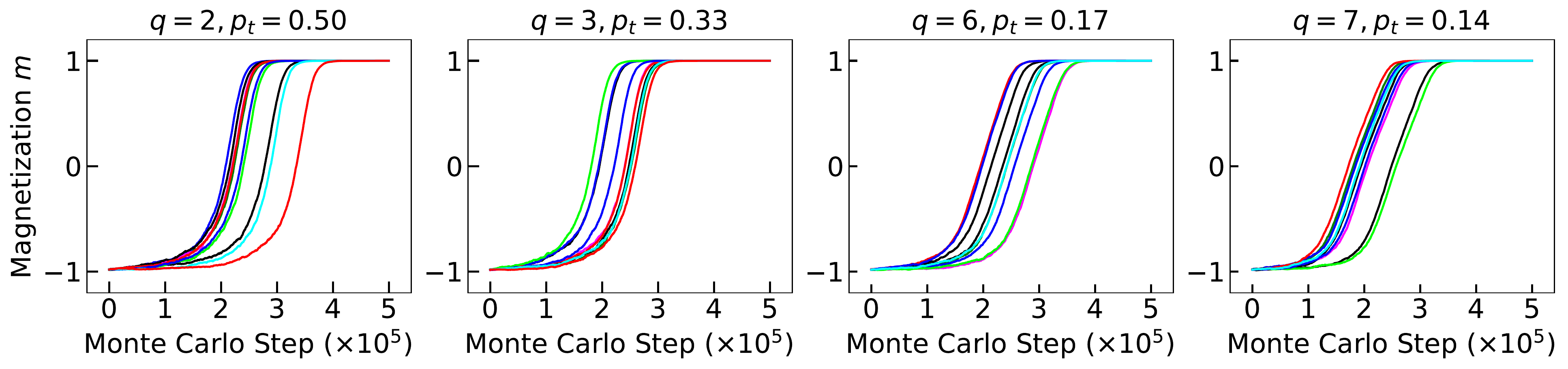}
\caption{\label{fig:Fig05} The evolution of the public opinion dynamics per site (represented by $m$) in time (Monte Carlo simulation step) of the model for a typical $q$-sized agent for $q = 2,3,6,7$. The population size $N = 10^4$ and the initial spin-up density $r = 0.01$. At the threshold probability $p_t$, all agents at the end have the same opinion `up' (magnetization $m = 1$),  indicating the {coexistence of two ordered states} has been eliminated at the probability threshold $p_t$.}
\end{figure}

To see more clearly the absence of the {coexistence of two ordered states} of the system at the probability threshold $p_t$, we simulate the evolution of order parameter $m$ for typical values of $q$-sized agents for the small of initial spin-up density $r = 0.01$ at probability threshold $p_t$ as exhibited in Fig.~\ref{fig:Fig05}. As seen, all samples reach an ordered state with magnetization $m = +1$ (complete consensus), indicating the absence of the {coexistence of two ordered states} for $p \geq p_t$. This result is consistent with the previous result (see Fig.~\ref{fig:Fig03}). In a social context, one can say that the effect of the mass media can lead the individual opinion to follow what the mass media wants. It has been shown that even though the initial opinion density is small (represented by the spin-up density $r$), with the influence of the mass media, community opinion becomes homogeneous (complete consensus) following mass media opinion.

\subsection{Time evolution}
We conveniently analytically formulated the time evolution of the spin-up density $r$ for any values of $q$-sized agent on the fully connected network. That is because, in the mean-field regime, the fluctuation of all agents is ignored; thus, the system's state can be represented only by one parameter, for instance, the density of spin-up $r$. In our previous work, we have discussed an agreement between the model on the fully connected network with the mean-field approximation in the statistical physics {for a large population size $N$ }\cite{MUSLIM2022133379, MUSLIM2022128307}. {The density of spin-up $r$ will either increase or decrease by $1/N$ or remain constant during the dynamics process, and the total probability of the process is one. We can write the probability transition of the spin-up increases symbolically as $\varphi(x-1/N \to x)$, decreases as $\varphi(x+1/N \to x)$, and remains constant as $\varphi(x \to x) =  1- [\varphi(x-1/N \to x)+ \varphi(x+1/N \to x)]$, and will be given explicitly in the next section. The rate equation of spin-up density $r$ can be derived from the discrete-time Master equation which corresponds to the time evolution of the probability distribution, after doing some simplifications \cite{jkedrzejewski2019statistical}. Thus, for a large population size $N$, the rate equation of spin-up density $r$ can be written as:} 
\begin{equation}\label{eq:recur_r}
    \dfrac{\mathrm{d}r}{\mathrm{d}t} =  \varphi^{+}(r)- \varphi^{-}(r).
\end{equation}

\begin{figure}[t]
\centering
\includegraphics[width = 6.5 cm]{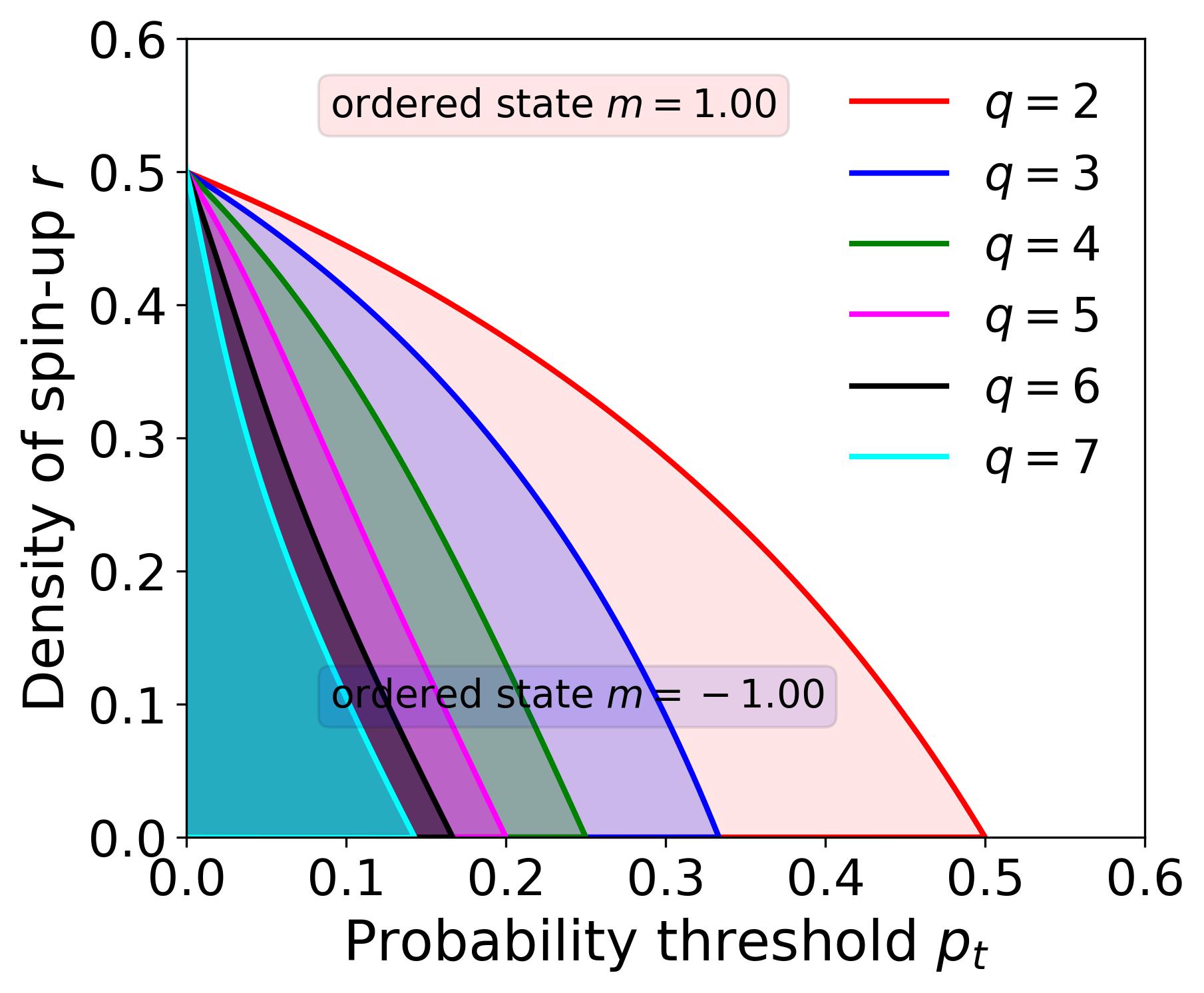}
\caption{\label{fig:Fig07} Phase diagram of the model [Eq.~\eqref{eq:critical}] for several values of $q$-sized agent. One can see that the upper and lower regions (colored region) of each $q$-sized agent are the ordered state with $r =  1\,(m =1)$ and $r = 0 \, (m=-1)$, respectively.}
\end{figure}

Based on the microscopic interactions we have mentioned earlier in the previous section, the probability spin-up increases $\varphi^{+}(r)$ and decreases $\varphi^{-}(r)$ for a finite system for the model on the fully connected network can be expressed as:
\begin{align}
    \varphi^{+} & = \dfrac{ N_{\downarrow}\prod_{i=1}^{q} \left(N_{\uparrow}-i+1\right)  }{\prod_{i=1}^{q+1}\left(N-i+1\right)}+ \dfrac{p\,N_{\downarrow}}{N} \left[1- \dfrac{ \prod_{i=1}^{q} \left(         N_{\uparrow}-i+1\right)-\prod_{i=1}^{q} \left(N_{\downarrow}-i+1\right)  }{\prod_{i=1}^{q}\left(N-i+1\right)}\right], \label{eq:vara}\\
     \varphi^{-} & = \dfrac{N_{\uparrow} \prod_{i=1}^{q} \left(N_{\downarrow}-i+1\right)}{\prod_{i=1}^{q+1}\left(N-i+1\right)}\label{eq:varb},
\end{align}
where $ N_{\uparrow}$ and $N_{\downarrow}$ are the total number of total spin-up and spin-down, respectively. {Note that Eqs.~\eqref{eq:vara} and \eqref{eq:varb} are the case for no repetition in the $q$-sized agent, given that the model is defined on a fully connected network identical to the mean-field model.} The second term in Eq.~\eqref{eq:vara} is a contribution from the existence of mass media effects that make agent opinion change from down to up opinions with probability $p$. Because the model is defined on the complete graph, we are more interested in considering models in the limit of large population $N$. Therefore, for a large $N$, Eqs.~\eqref{eq:vara} and \eqref{eq:varb} reduce to more simple forms such as:
\begin{equation}\label{eq:probabiliy_density}
    \begin{aligned}
    & \varphi^{+}(r) \approx r^q\, (1-r)+ p\,(1-r)\,\left[1-r^q-(1-r)^q\right], \\
    & \varphi^{-}(r) \approx r\, (1-r)^q.    
    \end{aligned}
\end{equation}
where $r = N_{\uparrow}/N$. We analyze the {separator} point $r_s$ of the model by considering the stationer condition of $r$ in Eq.~\eqref{eq:recur_r}, that is, $\varphi^{+} = \varphi^{-}$. One can see that obtaining the explicit expression of $r$ in $p$ for any values of $q$ will be difficult to do. However, we can express $p$ in $r$ for any values of $q$ and plotting it in a reserved way. For the stationary condition, we obtain:
\begin{equation}\label{eq:critical}
    p_t(r) = \dfrac{r^{q+1}+r\,\left(1-r\right)^{q}-r^{q}}{r^{q+1}-\left(1-r\right)^{q+1}-r^{q}-r+1}.
\end{equation}
Eq.~\eqref{eq:critical} also can be treated as the probability threshold of the mass media effect for any value of $q$, i.e., we will find that the system reaches a homogeneous state (complete consensus) for $p \geq p_t$. The plot of Eq.~\eqref{eq:critical} for a small $r = 0.01$ is shown in Fig.~\ref{fig:Fig04}. One can see the agreement between the analytical and the Monte Carlo simulation discussed in the previous section. We can also obtain the phase diagram from Eq.~\eqref{eq:critical} for any values of $q$-sized agent as shown in Fig.~\ref{fig:Fig07}. The phase diagram divides two areas where the ordered states reside. Those ordered states are the ones with $m = -1$ (colored) and those with $m = + 1$. In other words, the colored area is where we can always observe the system to reach a consensus state with $ m =-1$. In contrast, the non-colored area is where we will always obtain the system to reach a consensus with $ m =+1$. At $p_t = 0$, the {separator} point $r_s = 0.5$, and the increase of the $q$-sized agent makes the probability threshold decrease for a fixed spin-up density $r$. This condition can be understood that the selection of the larger $q$-sized agent in a random state impacts the probability of the system reaching a homogeneous state and becomes smaller exponentially as shown in Fig.~\ref{fig:Fig04} for the case with a small $r=0.01$.

\begin{figure*}[tb]
\centering
\includegraphics[width = 16 cm]{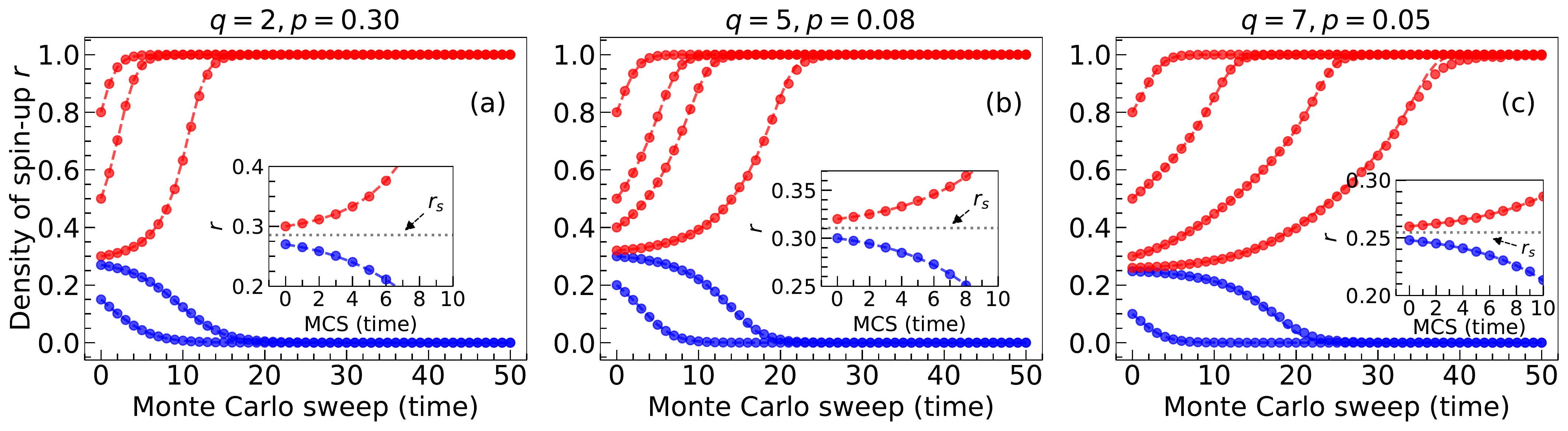}
\caption{\label{fig:Fig08} The comparison of the Monte Carlo simulation (data point) versus numerical calculation using R-K 4th order (dashed line) of the time evolution of  $r$ in Eq.~\eqref{eq:recur_r} of the model for typical values of $q$-sized agent, $q = 2$ [panel (a)], $q = 5$ [panel (b)], and $q = 7$ [panel (c)], and several values of  $p$ and $r$. One can see the agreement results of both methods. The blue color of the data for $r < r_s$, where all the data reach an ordered state with $m = -1$, and the red color for $r > r_s$, where all the data reach an ordered state with $m = 1$. The inset graph of each panel exhibits the data near the {separator} point $r_s$. The population size $N = 10^5$, and each point data averages $10^4$ independent realizations.}
\end{figure*}

The dependence of the spin-up density $r$ with time $t$ can be obtained from Eq.~\eqref{eq:recur_r}. The spin-up density $r$ can be theoretically computed for infinite systems such as $\int_{r_0}^{r} dr/\varphi(r) = \int_{t_0}^{t} dt$, but in such a way takes much work to carry out for any values of $q$. Instead, we can compute the evolution of $r$ using another numerical approximation such as the Runge-Kutta (R-K) fourth-order method \cite{pinder2018numerical}. However, for an illustration, we can obtain the exact solution of Eq.~\eqref{eq:recur_r} for a simple case, namely for $q = 2$ and without mass media effect $p = 0$, $r(t)_{1,2} = \Big[\left(r_0-1/2\right) \big[\left(r_0-1/2\right) \exp(t) \pm\big[\big(\left(r_0-1/2\right)^2 \exp(t)-r_0^2 +r_0\big)\exp(t)\big]^{1/2}\big]-r_0^2+r_0 \Big]/2\big[(r_0-1/2)^2 \exp(t)-r_0^2+r_0 \big]$, where $r(t)_1$ and $r(t)_2$ for $r_0 > r_s$ and $r_0 < r_s$, respectively. We can check that at $r_0 = 1/2$, the spin-up density $r(t)_{1,2}$ constant at $r_s = 1/2$ for all time $t$, while $r(t)_1 \to 1$ and $r(t)_2 \to 0$  for $t \to \infty$, characterized by the {separator} point at $r = 1/2$. The comparison between the Monte Carlo simulation versus the R-K results for $r$ for typical values of $q$-sized agent and probability $p$, $q = 2\,(p=0.3), 5\, (p = 0.08) $ and $q = 7\, (p =0.05)$ is shown in Fig.~\ref{fig:Fig08}. One can see the good agreement results, with the dashed line for the R-K method and the data point for the Monte Carlo simulation. The red color for the data with initial spin-up density $r$ is greater than the {separator} point ($r > r_s$ ), while the blue color for the data with $r < r_s$. All the data evolve to the stable point at $r = 0$ for $r_0 < r_s$ and $r = 1$ for $r_0 >r_s$. The inset figures visualize the evolution of $r$ close to $r_s$, where for $r > r_s$ and $r < r_s$, the system reaches a consensus with $m = +1$ and $m = -1$, respectively.

\subsection{Analyzing {separator} point from effective potential}
The unstable points that represent the {separator} point of the models can be analyzed using the effective potential $V(r)$ that can be expressed as:
\begin{equation}\label{eq:effective_potential}
    V(r) = -\int f(r)\, \mathrm{d}r,
\end{equation}
which is also called the unstable potential of the system. Potential in Eq.~\eqref{eq:effective_potential} can be used to analyze the movement of opinions within the system \cite{nyczka2012opinion}. In this section, we also analyze the opinion evolution within the system's potential framework to identify the instability point, which is the {separator} point separating the two ordered states. The term $f(r) = \varphi^{+}-\varphi^{-}$ is the ``effective force'' which drives the spins to flip during the dynamics process. In this section, because we consider a large  $N$, we use Eq.~\eqref{eq:probabiliy_density}
that expresses the probability for the spin-up density to increase $\varphi^{+}$ and decrease $\varphi^{-}$. Therefore, by inserting Eq.~\eqref{eq:probabiliy_density} into \eqref{eq:effective_potential}, and integrating it, the general solution of $V(r)$ for any values of $q$-sized agent is:
\begin{align}\label{eq:pot}
    V(r) = &\dfrac{\left(1 - p\right) r^{q + 1}}{\left(q + 1\right)\left(q + 2\right)}\left(r - 2 - \left(1 - r\right)q\right) - \dfrac{\left(1 - r\right)^{q + 1}}{\left(q + 1\right)\left(q + 2\right)}\left(1 + \left(q + 1\right)r\right) + \dfrac{p\,r\left(r - 2\right)}{2} + \dfrac{p \left(r-1\right)^2 \left(1 - r\right)^q}{\left(q + 2\right)}.
\end{align}

\begin{figure*}[tb]
\centering
\includegraphics[width = 16 cm]{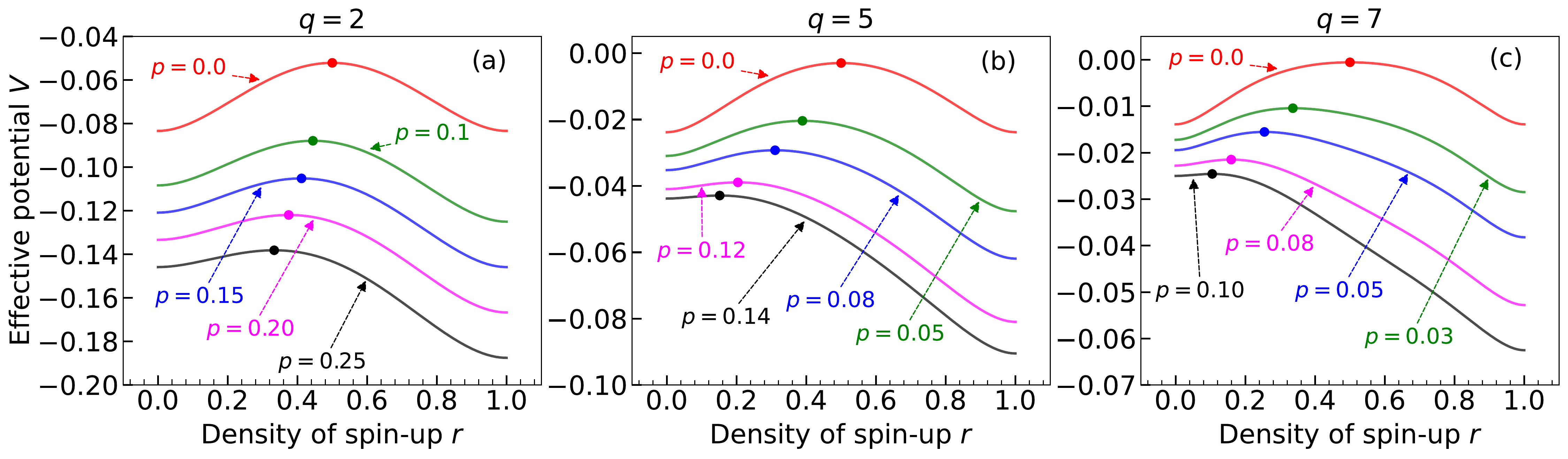}
\caption{\label{fig:Fig09} Effective potential in Eq.~\eqref{eq:pot} for $q = 2$ [panel (a)], $q = 5$ [panel (b)], and $q = 7$ [panel (c)], and typical values of  $p$. One can see the unstable point for each $p$ (top dot), indicating the {separator} point of each $p$. The increasing of $p$ makes the effective potential $V$ getting minimum and towards monostable at $ r = 1$, indicating that all spins reach an ordered state with the order parameter $m =  1$.}
\end{figure*}

Plot of Eq.~\eqref{eq:pot} for $q = 2, 5$ and $7$ for typical values of  $p$ can be seen in Fig.~\ref{fig:Fig09}. For $p$ increases, the effective potential $V(r)$ gets more minimum. The dot point of each probability $p$ is the unstable point ({separator} point) of the model that separates two stable points at $r = 0$ and $r =  1$. 

As we can see in Fig.~\ref{fig:Fig09}, for $p = 0.0$, the potential is symmetric to the spin-up density $r$, and the highest value at the {separator} point $r = 0.5$. In contrast, the stability points at $r = 0.0$ and $r = 1.0$ with the same value of $V$, indicating that the probability of reaching the complete consensus states with $ r = 0 \, (m = -1)$ and $r = 1 \, (m = 1)$ are the same. For $p \neq 0$, the unstable point shifts to the lowest $r$ as $p$ increases and the more stability point at $ r = 1$. We also see no transition order-disorder phase from the potential in this model. It is easy to understand that the existence of a probability $p$ does not act as noise, for example, the temperature in the Ising model, which destroys order, but rather makes the system achieve an ordered state. Therefore, the disordered state will never be reached in this model during the dynamics process, so it will never undergo an order-disorder phase transition. This condition can also be seen by plotting $r$ or $m$ vs. $p$ from Eq.~\eqref{eq:critical} [Fig.~\ref{fig:Fig07}]. We will see that when $p = 0$, $r = 1/2$ or $m = 0$ is at the equilibrium state instead of $r = 1$ or $m = 1$.

\subsection{Relaxation time and exit probability}

The other interesting parameters to discuss based on the social dynamics point of view are the relaxation time $\tau$ and the exit probability $E$. The relaxation time, also called the consensus time, is the duration the system needs to reach a complete consensus or an ordered state (all agents sharing the same opinion). The exit probability is defined as a probability that the system in the final state is in an up-state (agents, at the end at complete consensus state, either have up or down opinion) \cite{castellano2009statistical}. We will analyze these two parameters from the Fokker-Planck equation, which describes a system's diffusion process from one state to another state \cite{gardiner1985handbook}. For a large system, the Fokker-Planck equation can be written as:
\begin{equation}\label{eq:FK}
    \dfrac{\partial P(r,t)}{\partial t} =-\dfrac{\partial}{\partial r} \left[\chi_1(r)P(r,t) \right]+\dfrac{1}{2} \dfrac{\partial^2}{\partial r^2} \left[ \chi_2(r) P(r,t)\right],
\end{equation}
where $P(r,t)$ is the probability to find the system at state $r$ and time $t$. Note that the Fokker-Planck equation in Eq.~\eqref{eq:FK} is derived from the master equation after doing some formalism \cite{gardiner1985handbook}. In fact, one can see Eq.~\eqref{eq:FK} is just a one-dimensional Fokker-Planck equation since the model is defined on the fully-connected network corresponding to the mean-field character in statistical physics. In other words, the spatial dimensions do not affect the final results. The parameters $\chi_1(r)$ and $\chi_2(r)$ are the  drive-like and diffusion-like coefficients depending on the considered model, and for this model, $\chi_1(r) =  (\varphi^{+} -\varphi^{-})$ and $\chi_2(r) = (\varphi^{+}+\varphi^{-})/N$ are defined as:
\begin{equation} \label{eq:drift_diff}
    \begin{aligned}
        \chi_1(r) = &  \left[ r^{q} + p \left(1-r^{q}-\left(1-r \right)^{q}\right)-r \left(1-r \right)^{q-1} \right] \left(1-r \right) \\
        \chi_2(r) = &  \left[r^{q} +p \left(1-r^{q}-\left(1-r \right)^{q}\right)+r \left(1-r \right)^{q-1}\right] \dfrac{\left(1-r \right)}{N}.
    \end{aligned}
\end{equation}
Eq.~\eqref{eq:FK} should guarantee the probability density function $P(r,t) = 0$ for $ 0 > r > 1$ since $r \in [0,1]$. We can easily check that the drive-like and diffusion-like coefficients in Eq.~\eqref{eq:drift_diff} will be zero at $r = 0$ and $r = 1$, $\chi_1(0) = \chi_1(1) = 0$, and $\chi_2(0) = \chi_2(1) = 0$, which ensure that the magnetization cannot be either less or greater than $0$ ($|m| > 0$).

\begin{figure*}[t]
\centering
\includegraphics[width = 16 cm]{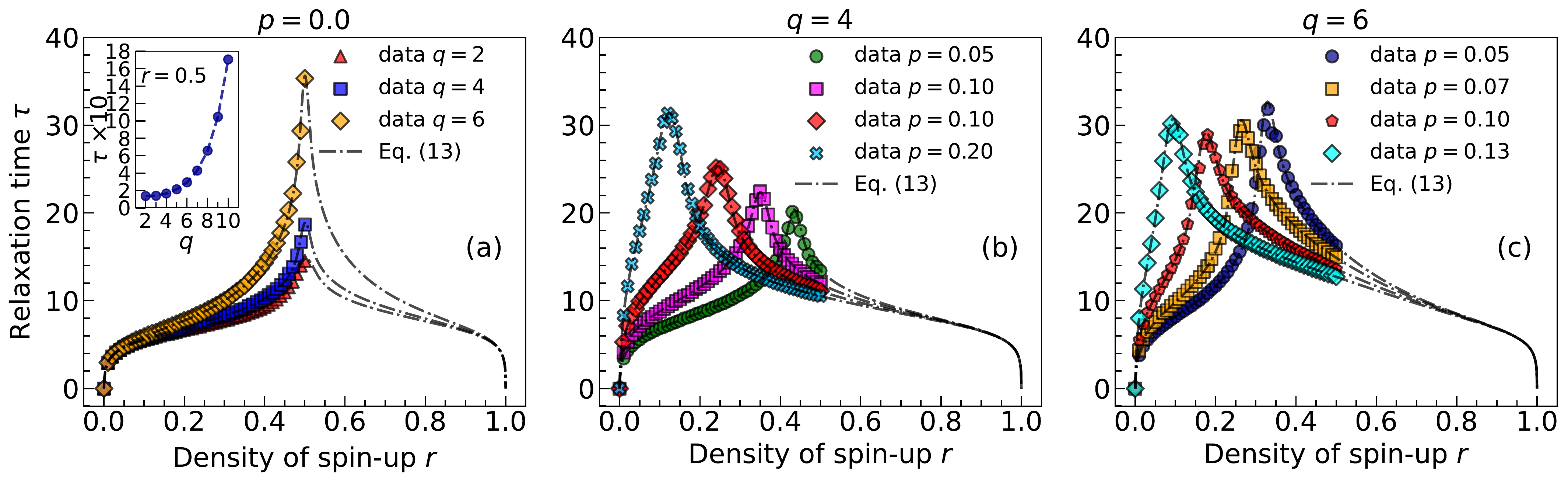}
\caption{\label{fig:Fig10} The comparison between Monte Carlo simulation versus analytical treatment [Eq.~\eqref{eq:relax_tau}] of the average relaxation time $\tau$ (in Monte Carlo sweep) for three cases, namely for without the mass media effect $p = 0.0$ [panel (a)], $q = 4$ [panel (b)], and $q = 6$ [panel (c)]. One can see the agreement results of both methods. The inset graph in panel (a) for the case with the most random state $r = 0.5$, namely the relaxation time $\tau$ increases for $q$ increases polynomially. The population size $N = 10^3$, and each data point averages $10^4$ independent realizations.}
\end{figure*}

We can generate the formulation of the relaxation time $\tau$ from the Fokker-Planck equation in Eq.~\eqref{eq:FK}; after doing some formalism, the differential equation of the relaxation time takes the form \cite{gardiner1985handbook}:
\begin{equation} \label{eq:relax_rel}
    \chi_1 \dfrac{\partial \tau (r)}{\partial r} + \dfrac{1}{2} \chi_2 \dfrac{\partial^2 \tau(r)}{\partial r^2}+1=0,
\end{equation}
with the boundary conditions $\tau(0) = \tau(1) = 0$, since there is no time is taken to make the system to a homogeneous state for the system with the initial state with all spin down ($r = 0$) and all spin up ($r = 1$). The general solution of  Eq.~\eqref{eq:relax_rel} is:
\begin{align} \label{eq:relax_tau}
\tau \left(r\right) = &
\int \Bigg[2 N \Bigg[\int \dfrac{\exp \Bigg[{-2 N \left(\int\dfrac{\left(\left(1-r \right) p +r \right) \left(1-r \right)^{q}+\left(1-p\right) (r^{1+q}+ r^{q})-p \left(1-r \right)}{\left(\left(1-r \right) p -r \right) \left(1-r \right)^{q}+\left(1-p\right) (r^{1+q}-r^{q})-p \left(1-r \right)} \mathrm{d}r \right)}\Bigg]}{\left(\left(1-r \right)p-r \right) \left(1-r \right)^{q}+\left(1-p\right) (r^{1+q}-r^{q})-p\left(1-r \right)}\mathrm{d}r \Bigg] \nonumber \\
& +C_1 \Bigg] \exp \Bigg[ {-2 N \left(\int\dfrac{\left(\left(1-r \right) p +r \right) \left(1-r \right)^{q}+\left(1-p\right) (r^{1+q}+r^{q})-p\left(1-r\right)}{\left(\left(1-r \right) p -r \right) \left(1-r \right)^{q}+\left(1-p\right) (r^{1+q}-r^{q})-p\left(1-r\right)}\mathrm{d}r \right)} \Bigg]\mathrm{d}r+ C_2.
\end{align}
where $C_1$ and $C_2$ are the constants integration that satisfies $\tau(0) = \tau(1) = 0$. One can see that the explicit form of $\tau(r)$ in Eq.~\eqref{eq:relax_tau} can be difficult to obtain exactly. However, we can solve Eq.~\eqref{eq:relax_tau} numerically \cite{gezerlis2020numerical}. The comparison results of the Monte-Carlo simulation (in Monte Carlo sweep) with Eq.~\eqref{eq:relax_tau} are exhibited in Fig.~\ref{fig:Fig10}. One can see the well-agreement result of both methods.

Without the mass media ($p = 0$), the relaxation time $\tau$ increases following the increase of the $q$-sized agent, as shown in Fig.~\ref{fig:Fig10} (a). This phenomenon is because the time to obtain a larger $ q$-sized agent in a homogeneous state is longer than for a smaller $q$-sized agent in a disordered state for the same spin-up density $r$. Therefore it affects the Monte Carlo step or sweep to reach a homogeneous state and also impacts the relaxation time of the system. One can also see in the change in spin-up density for every $q$ that the most random state of the system ($r = 0.5$) has the most significant relaxation time. The inset graph of Fig.~\ref{fig:Fig10} (a) shows the relaxation time $\tau$ for some values of $q$, and for $r = 0.5$ increases polynomially as $q$ increases. Panel (b) and (c) also show the agreement result of the Monte Carlo simulation versus Eq.~\eqref{eq:relax_tau}, where for $q = 4$ [panel (b)], the maximum relaxation time 
(peak) shifts to the lowest spin-up density $r$ and getting higher for {$p$ increases}. Panel (c) shows slightly different results from panel (b), {where the maximum relaxation time decreases for $p = 0.05, 0.07, 0.10$ and increases at $p = 0.13$}. The behavioral difference of the two values of $q$ can be due to the stochastic character of both $q$, where in the case of an order-disorder phase transition, the system can undergo a continuous (second-order) phase transition for $1 < q \leq 5$ and discontinuous (first-order) for $ q \geq 6$ for an independence noise character  \cite{nyczka2012phase, nyczka2013anticonformity}. The distribution of Monte Carlo sweeps corresponds to the relaxation time $\tau$ that follows a normal one for the case without mass media effect ($p = 0$) and a log-normal one for the case with mass media effect (not shown) in the region of the {coexistence of two ordered states} $p < p_t$. These distributions are similar to the impact of mass media introduced in the two-dimensional Sznajd model \cite{crokidakis2012effects}.

\begin{figure*}[t]
\centering
\includegraphics[width = 15 cm]{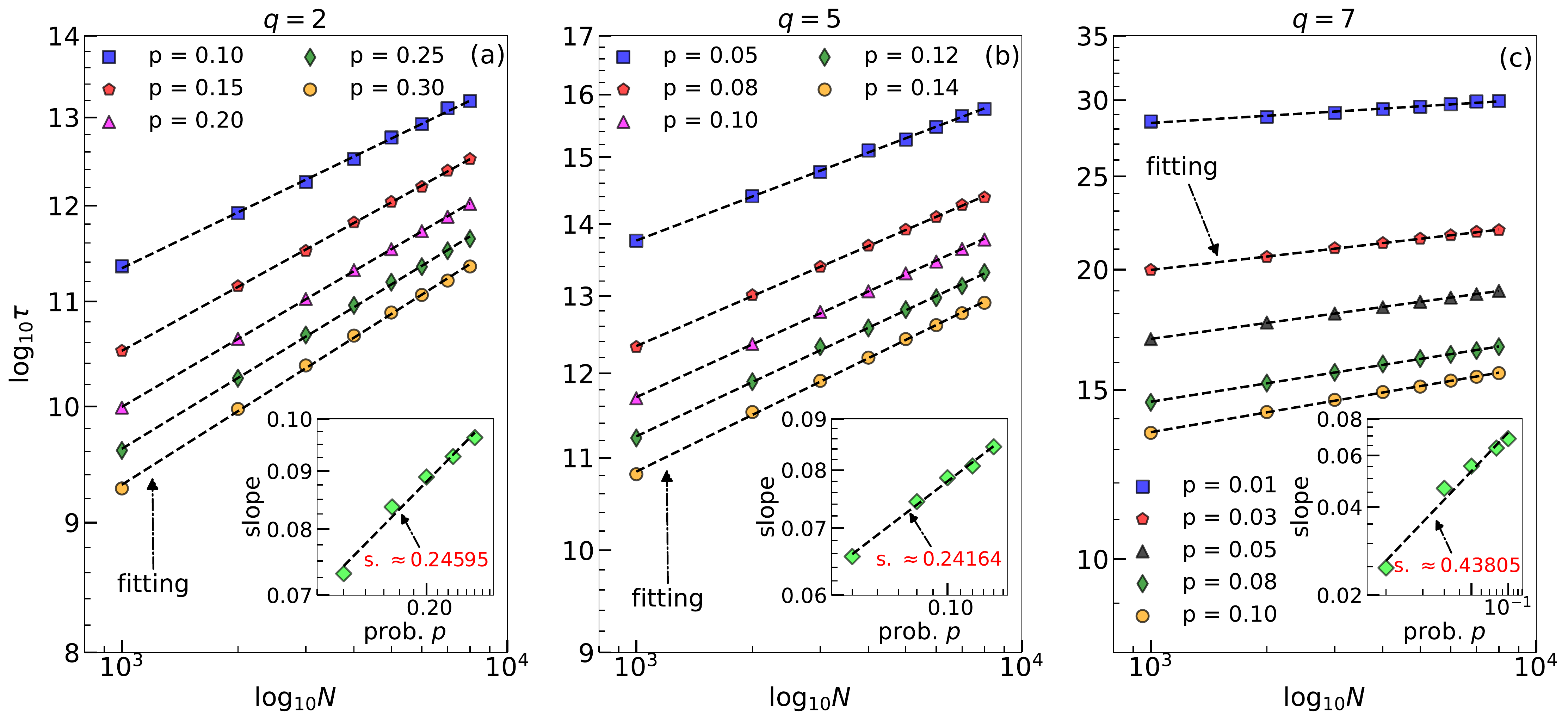}
\caption{\label{fig:Fig11} Log-log plot of  the relaxation time $\tau$ (in Monte Carlo step) versus the population size $N$ for typical values of $p$ for $q = 2$ [panel (a)], $q = 5$ [panel (b)], and $q = 7$ [panel (c)]. The relaxation time increases exponentially as $N$ increases with different slope values for each $p$. The inset graph shows that the slope increases as $p$ increases. The dashed line best fits the data, and each data point of $\tau$ averages $10^4$ independent realizations.}
\end{figure*}

We can also analyze the behavior of the relaxation time $\tau$ to the population size $N$. Fig.~\ref{fig:Fig11} (main graph) shows the log-log plot of the result of the Monte Carlo simulation of the relaxation time $\tau$ versus the population size $N$ for the spin-up density $r = 0.5$ and typical values of $q$-sized agent. As shown, the relaxation $\tau$ increases exponentially as $N$ increases for all probability values $p$. The slope of each data is not the same but increases as $p$ increases (depends on the probability $p$) as shown in the inset of Fig.~\ref{fig:Fig11}. Mathematically, we can write the relaxation time for a probability $p$ as $\tau \sim N^{\delta}$, where $\delta$ is the slope depends on the probability $p$ and $q$-sized agent. 

\begin{figure*}[t]
\centering
\includegraphics[width = 16 cm]{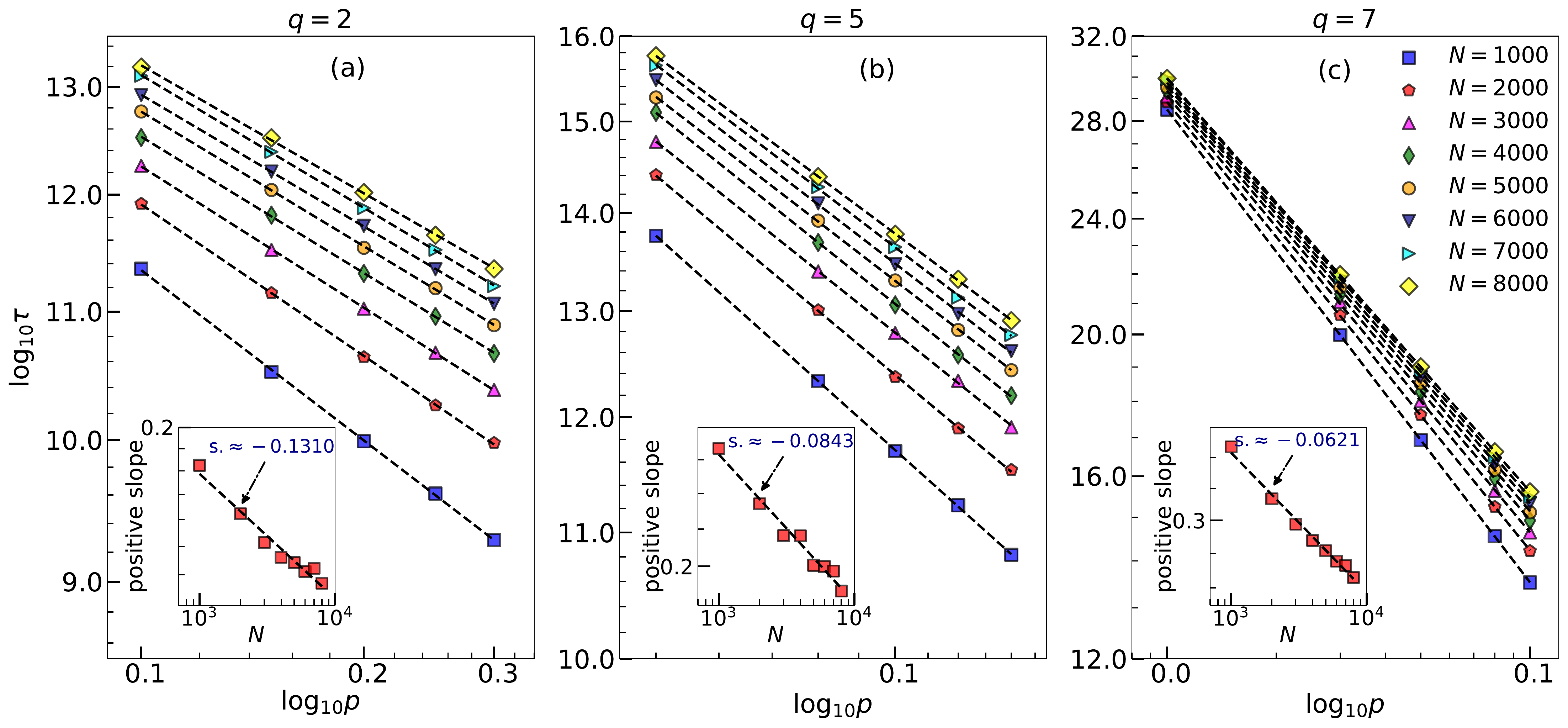}
\caption{\label{fig:fig_11} Log-log plot of the relaxation time $\tau$ (in Monte Carlo step) versus the probability $p$  for typical values of $p$ for $q = 2$ [panel (a)], $q = 5$ [panel (b)], and $q = 7$ [panel (c)]. One can see that the effect of the probability $p$ decreases the relaxation time $\tau$. The inset graphs show the log-log plot of the positive slope of each data $N$.}
\end{figure*}

Fig.~\ref{fig:fig_11} exhibits the log-log plot of the relaxation time $\tau$ versus probability $p$ for several values of $N$. The relaxation time $\tau$ decreases exponentially as $p$ increases. The inset graph exhibits the log-log plot of the positive slope of each data $N$, and the best fitting of all the data $q = 2, \, q = 5$, and $q = 7$ are $\eta \approx 0.1310, 0.0843,$ and $0.0621$, respectively. In this case, the relaxation time follows the power-law relation $\tau \sim N^{\eta}$, where $\eta$ depends on the $q$-sized agent with a power-law relation.

We also can generate the exit probability, that is, the probability of the system in a state up. The differential equation of the exit probability can be written as follows:
\begin{equation} \label{eq:exit}
    \chi_1 \dfrac{\partial E(r)}{\partial r} + \dfrac{1}{2}\chi_2 \dfrac{\partial^2 E(r)}{\partial r^2}=0,
\end{equation}
The general solution of Eq.~\eqref{eq:exit} for the exit probability $E(r)$ is:
 \begin{equation}\label{eq:exit_int}
    E\left(r\right) = C_1 + C_2 \left[
{\int}{\exp} \left[{-2 N \left({\int}\dfrac{\left(\left(1-r\right)p+r\right) \left(1-r \right)^{q}+\left(1-p\right) (r^{q +1}+ r^{q})-p \left(1-r \right)}{\left(\left(1-r \right)p+r \right) \left(1-r \right)^{q}+\left(1-p\right) (r^{q +1}-r^{q})-p \left(1-r \right)}\mathrm{d}r \right)}\right]\mathrm{d}r \right],
\end{equation}   

with boundary conditions $E(0) = 0$ and $E(1) = 1$. Similar to Eq.~\eqref{eq:relax_tau}, obtaining the exact solution of the exit probability $E(r)$ for any $p$ and $q$ can be difficult to do. Therefore we also solve Eq.~\eqref{eq:exit} numerically. However for a simple case i.e., for $q =2$ and $p = 0$, the solution of Eq.~\eqref{eq:exit_int} is $E(r) = 1/2\big[1+\erf{(\sqrt{N/2}\,(2r-1))}/\erf{(\sqrt{N/2}})\big]$. As expected, for $p = 0$, the exit probability $E(r)$ will coincide at $r = 1/2$ for any values of $N$, characterized by the {separator} point at $r_s = 1/2$. It is easy to understand that when there are no external influences (mass media) and the initial condition of the system is completely in a disordered state (the concentrations spin up and down are the same), then the probability of the system reaching a homogeneous state is the same, namely $1/2$.

\begin{figure*}[t]
\centering
\includegraphics[width = 16cm]{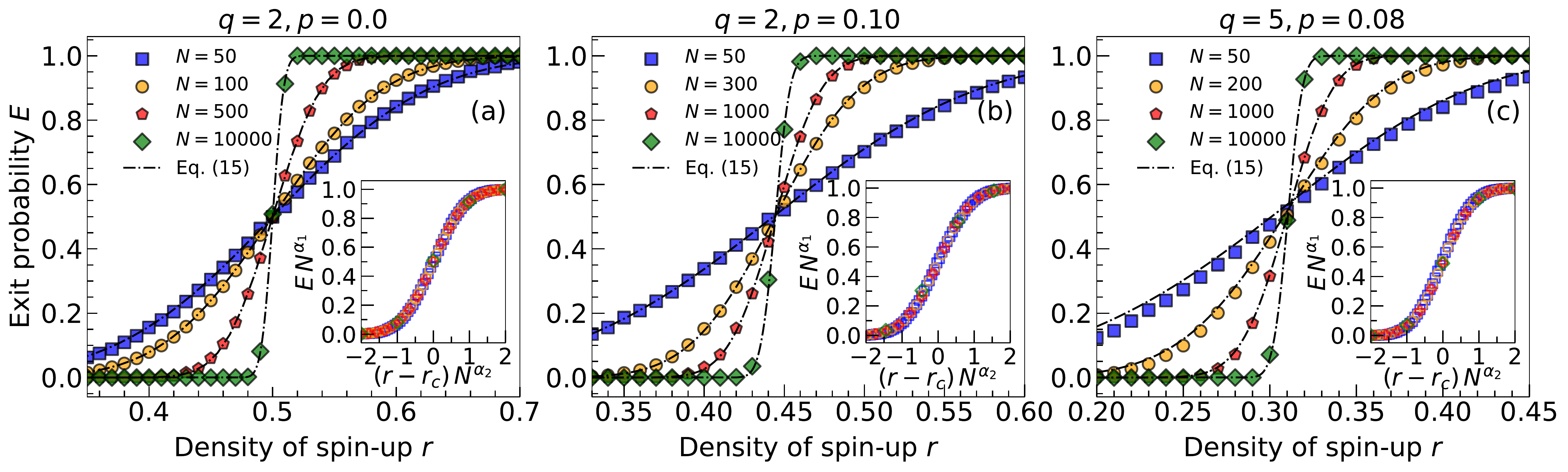}
\caption{\label{fig.012} Exit probability of the model on the fully-connected network for several values of number population $N$ and typical values of $p = 0, q = 0$ [panel (a)], $p = 0.10, q = 2$ [panel (b)], and $q = 5, p =0.08$ [panel (c)]. One can see the well-agreement results between the Monte Carlo simulation (data points) and Eq.~\eqref{eq:exit_int} obtained using numerical integration (dash-dotted lines). The critical {separator} point which separates two ordered states is obtained at the crossing lines of $E$, i.e., $r_s \approx 0.5$ for panel (a), $r_s \approx 0.445  $ for panel (b), and $r_s \approx 0.310$ for panel (c). The inset graphs show the scaling plot of the data, and obtaining the best scaling parameter that makes all data collapses are $\alpha_1 \approx 0.0 $, and $\alpha_2 \approx 0.5$ for all data. Each data point averages over $10^3$ independent realizations.}
\end{figure*}

We utilize a standard finite-size analysis to obtain the {separator} point of the model. The standard finite-size relation can be written as \cite{sousa2008effects}:
\begin{align}
     E(r,N)  & = N^{-\alpha_1}  f((r-r_s)\,N^{\alpha_2}), \label{eq:eq2}\\
     r(N) & = r_s+ \alpha_1 N^{-\alpha_2}, \label{eq:eq3}
\end{align}
where $\alpha_1$ and $\alpha_2$ can be referred to as the scale parameters that make the best fit for all the data. We investigate the range where the {coexistence of two ordered states} exists at $p < p_t$. The comparison between Monte Carlo simulation versus Eq.~\eqref{eq:exit_int} is exhibited in Fig.~\ref{fig.012}. One can see the well-agreement result between the Monte Carlo simulation and the analytical interpretation in Eq.~\eqref{eq:exit_int}, which is solved numerically. We obtain the {separator} point $r_s  \approx 0.5, 0.445$ and $0.254$ for ($q=2, p=0$), ($q = 2, p = 0.1 $), and ($q = 5, p = 0.08$), respectively as the result of the standard finite-size analysis in Eqs.~\eqref{eq:eq2}-\eqref{eq:eq3}. This {separator} point divides two ordered states with $m = +1$ for $r > r_s$ and $m = -1$ for $r < r_s$. The {separator} point, also called the unstable point, can be analyzed by directly investigating the crossing lines of exit probability $E$. The scaling parameters of the system that make the best fit for all data are $\alpha_1 \approx 0$ and $\alpha_2 \approx 0.5$ as shown in the inset of Fig.~\ref{fig.012}. These scaling parameters are universal; we can find the same $\alpha_1 $ and $\alpha_2 $ for all values of $N$. Moreover, these scaling parameters work for all values of $q$-sized agents in the region of existence of {two ordered states}.  We can also obtain the same scaling parameters $\alpha_1 \approx 0$ and $\alpha_2 \approx 0.5$ on other agent scenarios defined on the fully connected network \cite{azhari2022mass}.

\section{Summary and outlook}
We study the effect of mass media on opinion evolution based on the nonlinear $q$-voter model on the fully connected network. A $q$-sized agent and a voter are picked randomly in the population and interact based on the $q$-voter model. The voter adopts the $q$-sized agent whenever the $q$-sized agent has the same opinion. Otherwise, the mass media influence the voter. The mass media effect is represented as a probability $p$ that the voter follows the mass media opinion whenever the $q$-sized agent is not in a homogeneous state. We consider the mass media effect that leads the voter to adopt up-opinion. In a socio-political system, this scenario can illustrate a situation where the mass media takes sides or supports a particular choice.

We perform a numerical and analytical calculation for some macroscopic parameters of the model and obtain the agreement result between both methods. Based on our result, the {separator} point of the system is no longer at the spin-up density $r = r_s = 0.5$, as in the model without the mass media effect, but shifted to another $r$ and depends on the probability $p$. The {separator} point divides two ordered states with the order parameters $m = 1$ and $m = -1$ (represents the complete consensus situation with all up opinions ($+1$) and all down opinions ($-1$)). In other words, the system reaches an ordered state or complete consensus with $m = -1$ and $m = +1$ for the spin-up density $r < r_s \, (r > r_s)$ in a large limit $N$. Based on our results, we find that the mass media eliminates the {coexistence of two ordered states} at a probability threshold $p_t$, that is, a probability that guarantees that the opinion of the system always reaches a state of complete consensus, even for a small initial of opinion ``up'' (represented by spin-up density $r = 0.01$). In other words, two ordered states exists at $p < p_t$. 

We also analyze the {separator} point $r_s$  and the scaling parameters of the model using the standard finite-size scaling in Eqs.~\eqref{eq:eq2}-\eqref{eq:eq3} from the exit probability $E$ of the system. The {separator} point $r_s$ for each $q$-sized agent can be obtained directly from the intersection of lines between the exit probability $E$ versus spin-up density $r$. The scaling parameters of the model that make the best fit for all data are $\alpha_1 \approx 0.0$ and $\alpha_2 \approx 0.5$. These scaling parameters are universal, i.e., we find the same value for all population size values $N$. These scaling parameters can define the model's universality, which is similar to the majority-rule model in the case of the universality class \cite{azhari2022mass}. The parameters $\alpha_1$ and $\alpha_2$ are robust to the $q$-sized agents and work in the region of the existence of the {coexistence of two ordered states} ($p < p_t$). We also analyze the model's {separator} point from the effective potential of the system, and our finding shows that the {separator} point is the unstable point of the effective potential of the system.

The threshold probability $p_t$ decreases as the $q$-sized agent increases and obeys the power-law relation $p_t \sim q^{\gamma}$, where $\gamma = -1.00 \pm 0.01$, is the best scaling parameter of the model. The probability threshold is obtained in the equilibrium condition, that is, the rate of the spin-up density $r$, $dr/dt = 0$. The scaling behavior of this model is similar to the scaling behavior of the voting distribution of the 1998 Brazilian election reported by Costa et al. \cite{costa1999scaling}. We also examine the consensus time $\tau$ of the model, the duration needed for all voters to share the same opinion. We find that the average consensus time $\tau$ increases as $q$-sized agent increases for $p = 0$. For $p \neq 0$ and $q = 4$, the maximum relaxation time is getting higher for $r$ is getting smaller, while for $q = 6$, the maximum relaxation time is getting lower and bigger again for $r$ is getting smaller. In addition, the relaxation time follows the power-law relation with the population size $N$, that is,$\tau \sim N^{\delta}$, where $\delta$ is the slope depends on the probability $p$ and $q$-sized agent. The relaxation time increases exponentially as $N$ increases and decreases exponentially for $p$ increases following a power-law relation. 

Finally, from the sociophysics point of view, our results suggest that the mass media enormously influences society, although the initial amount of spin-up states is very small. It can lead the people's opinions to reach a complete consensus. It will be more interesting if this model is reviewed in several complex networks such as scale-free, small-world, Barabasi-Albert, and others.

\section*{CRediT authorship contribution statement}
\textbf{R.~Muslim:} main contributor, conducted analytical calculations and numerical simulations, visualization, \textbf{R.~Anugraha.~NQZ:} designed and supervised the research, \textbf{M.~A.~Khalif:} conducted numerical calculations. All authors reviewed and edited the manuscript.

\section*{Declaration of Interests}
The authors declare that they have no known competing financial interests or personal relationships that could have appeared to influence the work reported in this paper.


\section*{Acknowledgments}
This work is partially supported by Universitas Gadjah Mada. R.~Muslim. thank Dr. A.~R.~T.~Nugraha and Dr.~E.~H.~Hasdeo for helpful discussions in conducting the numerical simulation and the mini-HPC provided by the Quantum Matter Theory group, BRIN Research Center for Quantum Physics.

\bibliographystyle{elsarticle-num}

\bibliography{cas-refs}

\vskip3pt
\end{document}